\documentclass{article}[15pt]

\usepackage{amsmath,epsfig,color,calc,rotating}
\usepackage{wrapfig}
\usepackage{ulem}

\begin{document}

\title{
Fractal Characterizations of MAX Statistical Distribution 
in Genetic Association Studies
\vspace{0.2in}
\author{
Wentian Li$^{1}$ and 
Yaning Yang$^{2}$  \\
{\small \sl 1. The Robert S. Boas Center for Genomics and Human Genetics,  The Feinstein Institute }\\
{\small \sl for Medical Research, North Shore LIJ Health System,
 Manhasset, 350 Community Drive, NY 11030, USA.}\\
{\small \sl 2. Department of Statistics and Finance, University of Science and Technology of China,
 Anhui 230026, Hefei, CHINA}\\
}
\date{}
}
\maketitle  
\markboth{\sl Li, Yang) }{\sl Li, Yang}

{\bf ABSTRACT}: Two non-integer parameters are defined for MAX statistics, 
which are maxima of $d$ simpler test statistics. The first 
parameter, $d_{MAX}$, is the fractional number of tests, representing the
equivalent numbers of independent tests in MAX. If the $d$ tests are 
dependent, $d_{MAX} < d$.  The second parameter is the 
fractional degrees of freedom $k$ of the chi-square distribution $\chi^2_k$ 
that fits the MAX null distribution. These two parameters, $d_{MAX}$ and $k$,
can be independently defined,  and $k$ can be non-integer even if $d_{MAX}$ is 
an integer. We illustrate these two parameters using the example of 
MAX2 and MAX3 statistics in genetic case-control studies. We speculate
that $k$ is related to the amount of ambiguity of the model inferred by 
the test. In the case-control genetic association, tests with low $k$ 
(e.g. $k=1$) are able to provide definitive information about the
disease model, as versus tests with high $k$ (e.g. $k=2$) that
are completely uncertain about the disease model.  Similar to Heisenberg's 
uncertain principle, the ability to infer disease model and 
the ability to detect significant association may not be simultaneously 
optimized, and $k$ seems to measure the level of their balance.

\newpage

\large

\section{Introduction}

\indent

Geometric objects with non-integer dimensions such as coastal lines, 
random walk trajectories, Koch snowflakes have been well known
\cite{mandelbrot}.  Besides the feature of self-similarity, an
important property of most fractals is their non-integer dimensionality. 
It is perhaps less known that non-integer or fractional parameter
values is also a valid concept in statistical distributions. The best
example is the fractional degrees of freedom ($df$). The $\chi^2$  (chi-square)
distribution concerns the sum of squares of standard normal (Gaussian) variables.
If $X_1$ and $X_2$ are two normally distributed variables with
zero mean and unit variance, $Y= X_1^2 +X_2^2$ is then distributed
as the $\chi^2$ with two ($k=2$) degrees of freedom, denoted by $\chi^2_{k=2}$.
The analytic expression of the probability density distribution of $\chi^2_{k}$ 
is known: $0.5^{k/2}/\Gamma(k/2) \cdot x^{k/2-1} exp{(-x/2)}$,
where $\Gamma$ is the Gamma function \cite{numrec}.  In this expression, 
there is no conceptual difficulty to extend an integer value 
of $k$ to non-integers.  However, since there is a specific meaning of $k$ in the original
definition of chi-square distribution, i.e., the number of standard
normal variables to be summed, one may wonder whether non-integer
degrees of freedom, though allowed, have any applications.

The chi-square distribution plays an essential role in genetic association
analysis, whose goal is to determine whether a genetic marker on a 
particular chromosome location is associated (correlated) with a 
human disease or presence/absence of a phenotype of interest
\cite{weir,sham,thomas}.  The simplest
genetic marker has two possible ``values" (alleles), written as
$a$ and $A$. Because half of the genetic material of a person
is from the father (F), and another half from the mother (M), a marker
configuration can be written as F$|$M. The two-allele marker has four
possible configurations: $a|a, a|A, A|a$, and $A|A$. If we cannot 
distinguish the parental origin of an allele easily, as in the
case with most technologies current in use, $A|a$ and $a|A$ are grouped into
one configuration, and the resulting three configurations (after 
dropping the vertical bar), $aa, aA, AA$, are called genotypes.

The most popular design for genetic association study right now
is the case-control design \cite{risch,cordell,hirsch,balding,li08}.
In this design, a group of patients (cases) and a group of 
disease-free normal persons (controls) are recruited, whose
DNA molecules extracted, and their genotype throughout the genome
(e.g. 10$^5-10^6$ markers on 23 chromosomes) are determined (``genotyped"). 
For a particular marker, the number of case (and control) samples 
with the $aa, aA, AA$ genotypes are counted. These six genotype 
counts are stored in a 2-by-3 contingency table, rows for
two disease status and columns for three genotypes. Many null hypothesis
can be tested, and a significant violation of the null is
used as evidence for genetic association between the marker and the
disease.  Exploration of the protein-coding genes near the marker
could provide further insight into the mechanism for the disease.

Establishing the null hypothesis is not as easy as first thought. One
obvious choice is to assume the three genotype frequencies to be unchanged
in the two (case and control) groups. If we use the Peason's chi-square
test (goodness-of-fit test), the null distribution of the test statistic
is $\chi^2_{k=2}$. The relation between the degree of freedom $k$
and the size of the contingency table is straightforward: $k$ is equal to the
number of rows minus 1 multiplied by the number of columns minus 1 \cite{agresti}.
On the other hand, if the allele $A$ ``dominates"
allele $a$, there is no difference between the $aa$ and $aA$
genotypes; and after combining the $aa$ and $aA$ columns, the original
2-by-3 table becomes a 2-by-2 table, and the test statistic
follows the $\chi^2_{k=1}$ null distribution. The similar collapses from
2-by-3 to 2-by-2 table could be carried out in several other ways,
corresponding to ``recessive", ``multiplicative", ``over-dominant", etc.
disease models, each has a $\chi^2_{k=1}$ null distribution for
the corresponding test statistic.

If the disease model is known, i.e., if we know the disease risk
given a genotype, one can easily choose the null hypothesis 
and a test so that deviation from the null could be detected.
Unfortunately, for many complex human diseases, due to the 
multiple genes nature and gene-environment interaction, the disease
model for a specific risk gene is largely unknown. To increase the
chance to detect the association signal under the situation of
model uncertainty, several tests, each testing a different null hypothesis, 
could be applied, and the best result among them is used. 
We call this procedure the ``MAX test". The MAX test that maximizes
the test statistics from two or three disease models is a
compromise between using one simple disease model and using
no models. As a result, the null distribution of MAX test statistics
is neither $\chi^2_{k=1}$ nor $\chi^2_{k=2}$, but something in
between.  We will show that this indeed leads to a fractional degrees of freedom
$k$ for $\chi^2_k$, and $2-k$ measures our knowledge about
the disease model. The determination of $k$ is complicated by
another issue that sometimes the two or three test statistics
being maximized are not independent. That leads to another
fractional parameter:  the number of independent tests $d_{MAX}$.

Although these two fractional parameters are not the same as 
the fractal dimension for fractals, a common theme is the non-integer
value. We will explore the properties of these two parameters
in details in this paper, organized as follows: Section 2 introduces statistical tests
and MAX statistical test; Section 3 discusses the fractional
number of tests for MAX test, from the perspective of family-wide $p$-values; 
Section 4 discusses the fractional degrees of freedom of chi-square
distribution, from the perspective of fitting null distribution
of MAX test statistics.  In the discussion section, we address
the issue on whether the fractional degrees of freedom is connected
to fractal dimension in the parameter space.

\section{MAX statistical test}

\indent

When two different statistical tests are carried out on the same 
dataset \cite{miller}, the more significant result of the two 
(i.e., the more extreme test statistic value) can be reported as 
the overall test result. This is a MAX statistical test. Clearly, 
MAX test statistic will always be larger than (or at least equal 
to) individual tests being maximized. Although the null distribution 
of a MAX test statistic may not be expressible by a simple 
analytic formula, we do expect the ``center of gravity" of 
the distribution to be shifted to the right to have a larger mean 
and a thicker tail area (``inflated type I error"), as compared 
to that of a individual test, for the obvious reason that the 
maximization procedure increases the mean value.

Here we would like to define a MAX statistic for the case-control
genetic association study. A dataset of such study consists
of six numbers: number of case samples with $aa, aA, AA$ genotypes 
($N_{10}, N_{11}, N_{12}$), and the number of control samples with these three 
genotypes ($N_{00}, N_{01}, N_{02}$) (see Appendix). 
The row $i$ in $N_{ij}$ indicates 
the case (1) or control(0) status, and column $j$ indicates the genotype 
with $j$ copies of $A$ allele. 

We consider three different tests which are part of a test family
called Cochran-Armitage trend ($CAT$) test \cite{cochran,armitage}.
This family of tests is parameterized by a $x$ value, and the null hypothesis is 
the equality of weighted genotype frequency $x P_{aA} +P_{AA}$ in case and control 
group. When $x=0$, we are testing $P_{AA, case}=P_{AA,control}$,
which corresponds to the genetic recessive model on the risk allele $A$. When $x=1$,
we are testing the equality of $P_{aA}+P_{AA}$ in case and control group,
which corresponds to genetic dominant model (whenever the risk allele $A$ is
present in a genotype, the disease risk is the same regardless 
of the second allele).
When $x=0.5$, we are testing the equality of the allele frequency,
$P_{aA}/2+ P_{AA}=P_A$, in the two groups.

The expression of $CAT$ test statistic is given in Appendix.
There are other reformulation of the above formula, such as
using the estimated allele frequency difference and Hardy-Weinberg
disequilibrium coefficient difference \cite{li-cbc}, but the
simplest calculation of $CAT(x=0)$ or $CAT(x=1)$ is to merge the
$aA$ counts ($j=1$) with the $aa$ counts ($j=0$) or $AA$ counts ($j=2$),
then calculate the Pearson's chi-square test statistic (see Appendix).

\begin{figure}[t]
  \begin{turn}{-90}
   \epsfig{file=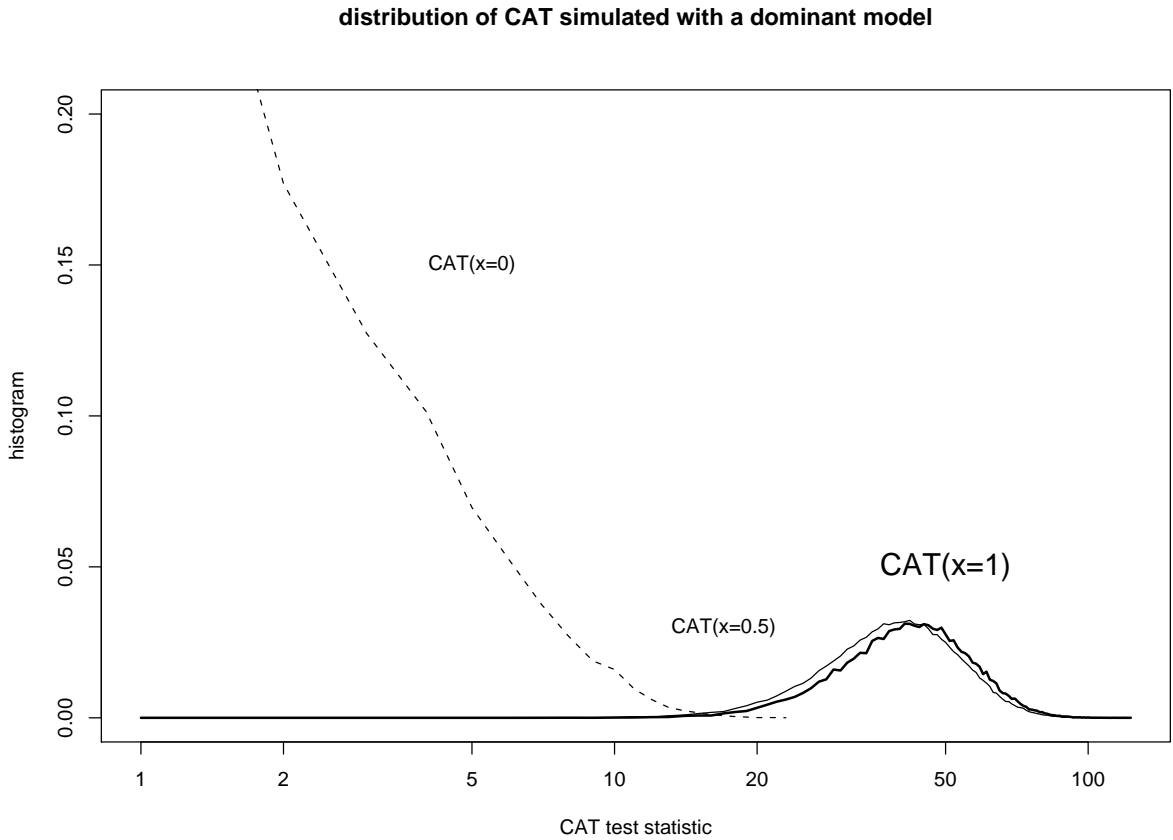, width=11cm}
  \end{turn}
\caption{
\label{fig1}
The distribution of $CAT(x=1)$, $CAT(x=0.5)$ and $CAT(x=1)$ from
the 100,000 replicates generated by a dominant model:
population risk allele frequency $p=0.1$, penetrance for 
baseline homozygote is 0.005, and genotype relative risk
for both heterozygote and the risk homozygote is $\lambda_1=\lambda_2=2$.
The genotype frequency for the case and the control group is
calculated by the formula given in \cite{wittke}.
}
\end{figure}

If the underlying disease model is dominant, multiplicative,
or recessive, the $CAT(x=1)$, $CAT(x=0.5)$, or $CAT(x=0)$,
respectively, tends to be the largest. Fig.\ref{fig1} shows an example
using a dominant model. The histogram determined by 100,000 replicates
for $CAT(x=1)$ is peaked at the higher value than the other two $CAT$'s,
$CAT(x=0.5)$ is distributed slightly lower than $CAT(x=1)$, whereas
the distribution of $CAT(x=0)$ is far towards the smaller values.

If the disease model is unknown, it is when a MAX statistic is useful.
One may consider these MAX statistics for case-control genetic data:
\begin{eqnarray}
\label{eq1}
MAX2 & \equiv & max(CAT(x=0), CAT(x=1)) \\ \nonumber
MAX3 & \equiv & max(CAT(x=0), CAT(x=0.5), CAT(x=1))  
\end{eqnarray}
MAX2 was discussed in \cite{suh,li-cbc}, and MAX3 was discussed in
\cite{freidlin,zheng03,zheng06}. Both MAX2 and MAX3 are ``smart" 
samplings of the disease model space without an exhaustive search.

\section{The fractional number of tests in calculating test-family wide $p$-values
in MAX statistics}

\indent

When several tests are applied to the same dataset and these
tests are independent, there is a simple formula for calculating
the test-family-wide $p$-value, which is also the $p$-value for 
the MAX test. We can derive the tail probabilities under the null
distribution (i.e., $p$-value), $p_{MAX2}$ and $p_{MAX3}$,  with the 
tail starting from the observed test statistic value $M$ by
the following procedure ($p_{\chi^2}$ is the tail area probability 
under $\chi^2_{k=1}$ distribution):
\begin{eqnarray}
\label{eq:sidak}
p_{MAX2} & \equiv & P( MAX2 > M |null) \nonumber \\
& = & 1- P( MAX2 < M |null) \nonumber \nonumber \\
 &=& 1- P( CAT(x=0) < M ~~ and ~~ CAT(x=1) < M |null)  \nonumber \\
 & \approx&  1- P( CAT(x=0) < M |null) \times P( CAT(x=1) < M |null)  \nonumber \\
&=& 1- (1-P( CAT(x=0) > M |null)) \times (1-P(CAT(x=1) > M |null )) \nonumber \\
 &=& 1- (1- p_{\chi^2})^2 \nonumber \\
p_{MAX3} & \equiv & P(MAX3 > M |null) \nonumber \\
  &=& 1- P( MAX3 < M |null) \approx \dots = 1-(1- p_{\chi^2})^3 .
\end{eqnarray}
The approximation  can be replaced by the equal sign for MAX2 only
if $CAT(x=0)$ and $CAT(x=1)$ are independent, and for MAX3 only if $CAT(x=0)$, $CAT(x=1)$ 
and $CAT(x=0.5)$ are independent. The independence assumption is untrue for 
MAX3 \cite{freidlin}, but close to be true for MAX2 \cite{li-cbc}. The two approximate formula
in Eq.(\ref{eq:sidak}), also known as Dunn-\u{S}id\'{a}k formula
\cite{ury,sokal}, can be written as $1- (1-p_{\chi^2})^d $ for $d$ 
tests being maximized in MAX.

If we force the approximation sign in Eq.(\ref{eq:sidak}) to be equality, 
$d$ can be derived from  $p_{MAX}$. This value of $d$ (called
$d_{MAX}$ here) represents the {\sl effective} number of independent tests:
\begin{equation}
\label{eq:dmax}
d_{MAX} = \frac{ \log (1- p_{MAX})}{ \log (1- p_{\chi^2})}.
\end{equation}
Note that the tail area probabilities for both MAX and chi-square, 
$p_{MAX}$ and $p_{\chi^2}$, are determined by the same $M$, the starting
position of the tail area.

Besides multiple testing correction in test-family-wide $p$-value on the same
dataset, the Dunn-\u{S}idak formula can also be used with 
{\sl the same test} on {\sl multiple datasets}. In particular, in whole genome 
association or linkage studies, selecting the SNP with the best 
association or linkage signal among $\sim 10^5$ SNPs belong to 
this application \cite{lander,efron04}, and the genome-wide $p$-value 
is calculated in the same way. The severe correction on $p$-value
in this application is in a sharp contrast to the correction 
in Eq.(\ref{eq:sidak}), of a factor of only 2 or 3.

In order to estimate the effective number of tests for
MAX2 and MAX3, we carried out the following simulation. 
We generated $N_r=$100,000 replicates, each replicate
is a 2-by-3 genotype counts for 1000 cases and 1000 controls.
The allele frequency is randomly chosen but the same
allele frequency is used to simulate both case and control
genotypes. The genotype frequency is derived from the allele
frequency by the Hardy-Weinberg equilibrium. The empirical 
distribution of MAX2, MAX3, $CAT(x=0)$, $CAT(x=1)$, 
$CAT(x=0.5)$ can all be determined with 100,000 realizations of
test statistic values (and the minimum $p$-value can't be
smaller than 1/100,000$=10^{-5}$).
Using several threshold $M$ value (controlling type I error),
$d_{MAX2}$  and $d_{MAX3}$ can be calculated by Eq.(\ref{eq:dmax}).
Two more runs were also carried out with 3000 cases/3000 controls,
and 5000 cases/5000 controls.

\begin{figure}[t]
  \begin{turn}{-90}
   \epsfig{file=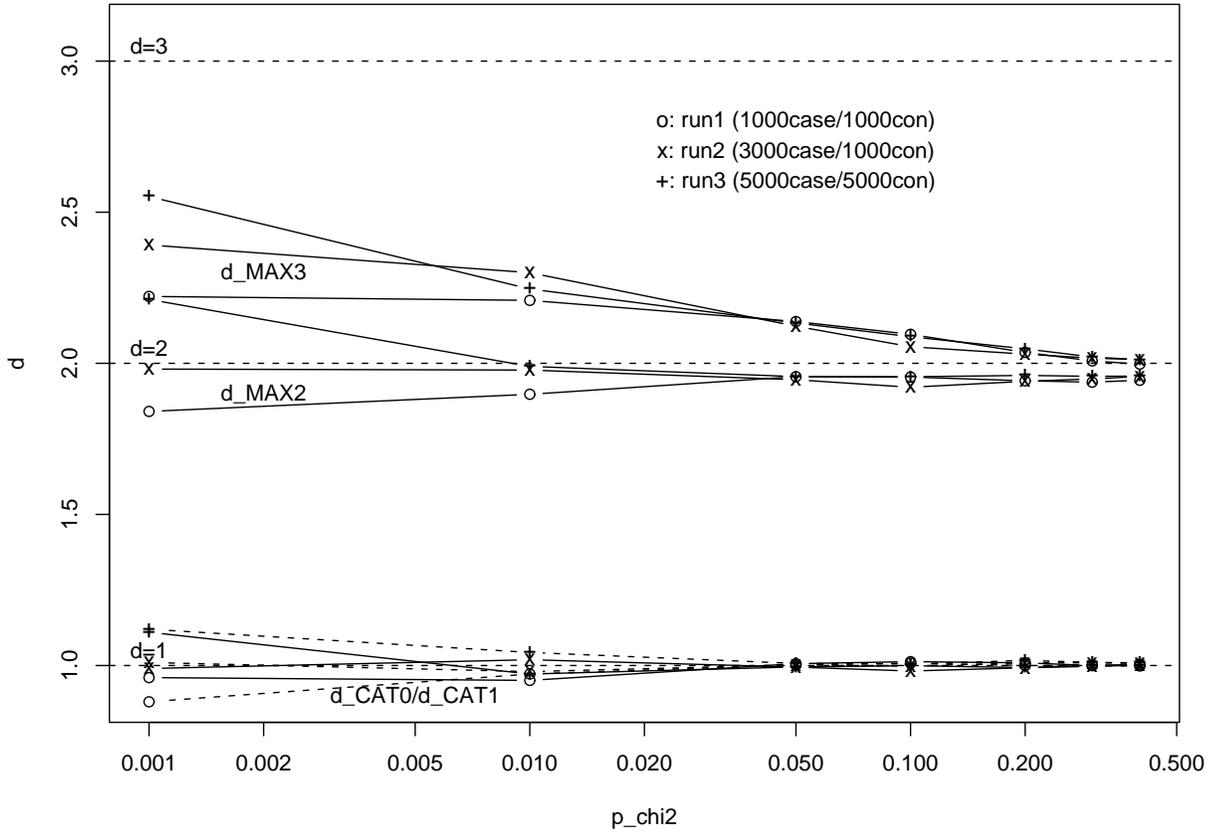, width=11cm}
  \end{turn}
\caption{
\label{fig2}
Fractional number of tests for MAX2 and MAX3 ($d_{MAX2}, d_{MAX3}$)
determined by Eq.(\ref{eq:dmax}) with three simulation runs, 
as a function of tail area probability under $\chi^2_1$ ($p_{\chi^2_1}$). 
Each run contains 100,000 replicates of genotype count 
tables for 1000 cases and 1000 controls (3000 cases/3000 
controls, 5000 cases/5000 controls for the second and 
the third run). As a comparison, the effective number of tests
for $CAT(x=0)$ and for $CAT(x=1)$ as determined by simulation is
also included. As expected, these effective number of tests is
essentially equal to 1.
}
\end{figure}

Fig.\ref{fig2} shows the empirical $d_{MAX2}$ and $d_{MAX3}$ 
as a function of $p_{\chi^2}$, the tail probability for the
$\chi^2_1$ distribution. 
It can be seen that although there is only a slight reduction
of $d_{MAX2}$ from the expected value of 2, $d_{MAX3}$ is
much smaller than the expected value of 3. At $p_{\chi^2}=0.05$,
the value of $d_{MAX3}$ is around 2.1, consistent with a similar 
result of $d_{MAX3}=2.2$ in \cite{gonzalez}.

The empirical $d$ values calculated from Eq.(\ref{eq:dmax}) 
for $CAT(x=1)$ and $CAT(x=0)$ are also shown 
in Fig.\ref{fig2} as a check of accuracy of the simulation. 
Indeed, the $d$ values do not deviate from the
expected value of 1 with the exception at smaller $p_{\chi^2}$ values.
For low $p_{\chi^2}$ values, a smaller number of replicates are
used in the determination of the empirical $p$-values, thus 
variance is large -- Fig.\ref{fig2} does show that
the estimated $d_{MAX2}$ and $d_{MAX3}$'s are not consistent
among the three runs at (e.g.) $p_{\chi^2} < 0.01$, an indication of
large run-to-run variation.  

Another source of potential 
bias is that  we keep the allele frequency a minimum distance away 
from the 0 value in order to avoid the situation of zero 
genotype count. The range of allele for these 3 runs
are (0.1, 0.9), (0.05, 0.95), and (0.02, 0.98) respectively.
Only when both the sample size and number of replicates
go to infinity, and with unconstrained allele frequency, 
can one expect the simulation-based estimation of $d_{MAX}$ 
values to be exact.

\section{Chi-square distributions with fractional degrees of freedom
that fit the null distribution of Max test statistics}

\indent

The second fractional parameter value related to the MAX concerns the
fitting of MAX null distribution by a non-integer-$k$ $\chi_k^2$ distribution.
As mentioned in Section 1, non-integer-$k$ chi-square distribution 
$\chi^2_{k}$ can be determined easily and is indeed implemented 
in statistical packages, such as $R$ ({\sl http://www.r-project.org/}).
Here we would like to check which $k$ value in $\chi_k^2$ leads to a 
better fit to the MAX null distribution.

In order to avoid confusion between $k$ and $d_{MAX}$, we made 
component tests to be independent so that $d_{MAX}$ remains 
an integer. Instead of generating case and control samples 
with specific genotype then calculate the MAX2, MAX3, $CAT(x=1)$ 
and $CAT(x=0)$, we randomly sample two, or three independent 
chi-square values from the $\chi^2_{k=1}$ distribution, then 
the maximization procedure is carried out. Due to the 
independence between chi-square values, $d_{MAX2}$ and $d_{MAX3}$
should be exactly equal to 2 or 3. Here we use a different
notation, Max2 and Max3, to represent this correlation-free simulation
(to be compared with Eq.(\ref{eq1}):
\begin{eqnarray}
Max2 &\equiv & max( \chi^2_1, \chi^2_1) \nonumber \\
Max3 &\equiv & max( \chi^2_1, \chi^2_1, \chi^2_1) 
\end{eqnarray}

\begin{figure}[t]
  \begin{turn}{-90}
   \epsfig{file=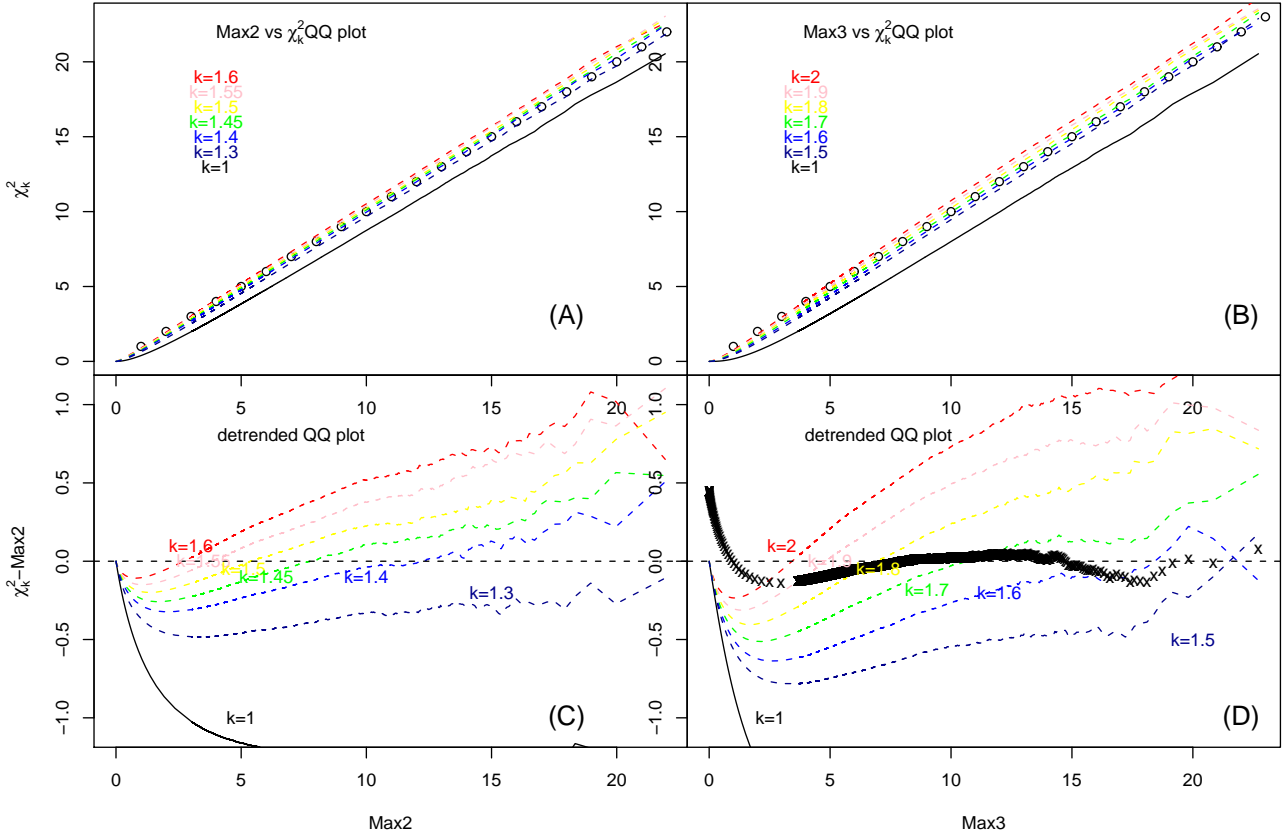, width=11cm}
  \end{turn}
\caption{
\label{fig3}
Quantile-quantile (QQ) plot of Max2/Max3 against values
sampled from $\chi^2_k$ with fractional degrees of freedom $k$. 
(A) QQ plot of Max2 against  values sampled from
$\chi^2_k$'s with $k=1, 1.3, 1.4, 1.45, 1.5, 1.55, 1.6$.
The circles indicate the QQ-plot between two identical
distributions. (B) QQ plot of Max3 against values sampled from 
$\chi^2_k$'s with $k=1, 1.3, 1.5, 1.6, 1.7, 1.8, 1.9, 2$. 
(C) Detrended QQ plot of Max2 against values sampled from
$\chi^2_k$'s. (D) Detrended QQ plot of Max3 against values
sampled from $\chi^2_k$'s. The crosses represent the detrended
QQ-plot for $0.45 +0.96 \chi^2_{k=1.7}$ against Max3.
}
\end{figure}

Fig.\ref{fig3} shows the result of fittings the empirical 
Max2 and Max3 by chi-square distribution with non-integer degree of freedoms.
Fig.\ref{fig3}(A,B) are the quantile-quantile (QQ) plot, where the
$x$-axis is the ranked Max2 or Max3 value and $y$-axis is the ranked
chi-square values with a fractional degrees of freedom 
($k$=1.3, 1.4, 1.45, 1.5, 1.55, 1.6 for Max2,  and $k$=1.5,
1.6, 1.7, 1.8, 1.9, 2 for Max3). To reduce variation, the average
of 100 runs is used in Fig.\ref{fig3}. When two distributions 
are identical, their QQ-plot should trace
the diagonal line with slope=1 (marked by circles). In Fig.\ref{fig3}(A,B)
chi-square distribution with a range of fractional degrees of freedom
seem to fit the Max2 and Max3 distribution well. 

To examine more carefully how good fractional $k$ chi-square 
distributions fit the Max2/Max3 distribution, we draw the detrended
QQ-plots in Fig.\ref{fig3}(C,D), i.e., $y$-axis is
the difference between the sorted chi-square values with fractional $k$ and
the sorted Max2 or Max3 values. Fig.\ref{fig3}(C,D) show 
systematic deviations between the two distributions. In other
words, no chi-square distribution with one single fractional $k$ value
may fit Max2 and Max3 for the entire range of values. For example, 
at Max2 $\approx 5$, $\chi^2_{1.5} -$ Max2$\approx 0$ (good fit),
whereas when Max2 $ >> 5$, $\chi^2_{1.5} -$ Max2$ >$0 (bad fit).

\begin{figure}[th]
  \begin{turn}{-90}
   \epsfig{file=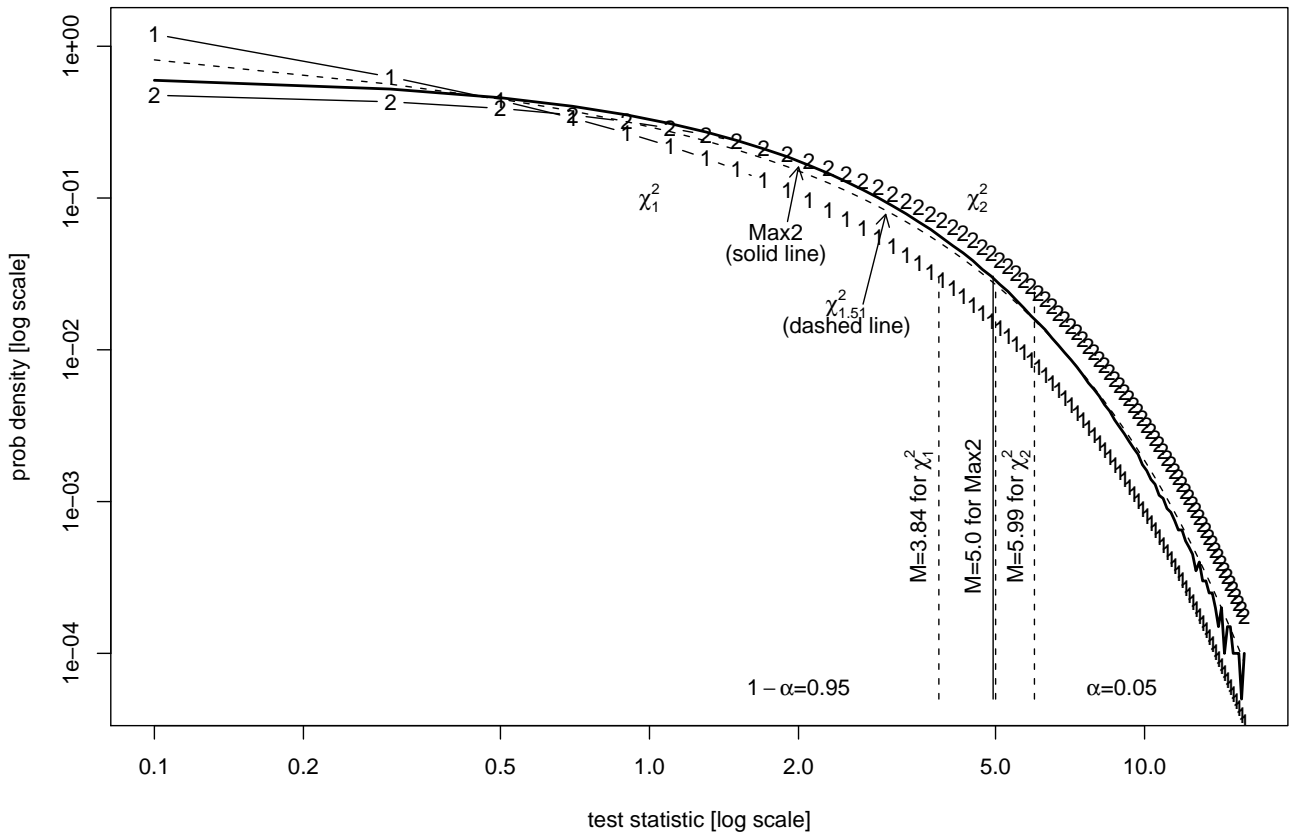, width=11cm}
  \end{turn}
\caption{
\label{fig4}
Probability density distribution of $\chi^2_1$ (labeled by 1),
$\chi^2_2$ (labeled by 2), $\chi^2_{k=1.51}$ (dashed line),
and simulated Max2 (solid line). The threshold value $M$'s
that correspond to tail area of 0.05 for these distributions
are also marked. The $\chi^2_{k=1.51}$ distribution has the
same $M$ as the Max2 distribution, so the two are equivalent
at the tail area of 0.05.
}
\end{figure}

It is straightforward to determine which $\chi^2_{k}$ crosses
the zero horizontal line at what position in Fig.\ref{fig3}(C,D).
First, from Eq.(\ref{eq:sidak}), we see a simple relationship 
between the ``head area" of $\chi^2_1$ and that of Max2/Max3:
\begin{eqnarray}
\sqrt{ 1-p_{Max2}} &=& 1-p_{\chi^2} \nonumber \\
(1-p_{Max3})^{1/3} &=& 1-p_{\chi^2}. 
\end{eqnarray}
The approximation in Eq.(\ref{eq:sidak}) becomes equality because 
MAX2/MAX3 is replaced by Max2/Max3. Here is an example in
determining the zero crossing point in Fig.\ref{fig3}(C): 
if the tail area $p_{Max2}$ for Max2 is 0.05, the ``head area" is 0.95,
and the corresponding head area for $\chi^2_1$ is $\sqrt{0.95}= 0.9746794$.
That head/tail area for  $\chi^2_1$ can be used to determine the
threshold value $M=5.001825$, as marked in Fig.\ref{fig4}.

Then, we choose a $\chi^2_k$ with fractional $k$ so that its
tail area determined by $M=5.001825$ is also 0.05. As shown in
Fig.\ref{fig4}, the threshold value for 0.05 area for $\chi^2_1$ 
is 3.841459 and that for $\chi^2_2$ is 5.991465. A fractional-$k$
$\chi^2_k$ ($1 < k < 2$) should have the threshold value for
0.05 tail area at $M=5.001825$. The exact $k$ can be iteratively
determined by a bisection method, resulting in $k=1.51$.
In other words, at $p_{Max2}=0.05$, $\chi^2_{k=1.51}$ is
equivalent to the null distribution of Max2.

\begin{figure}[th]
  \begin{turn}{-90}
   \epsfig{file=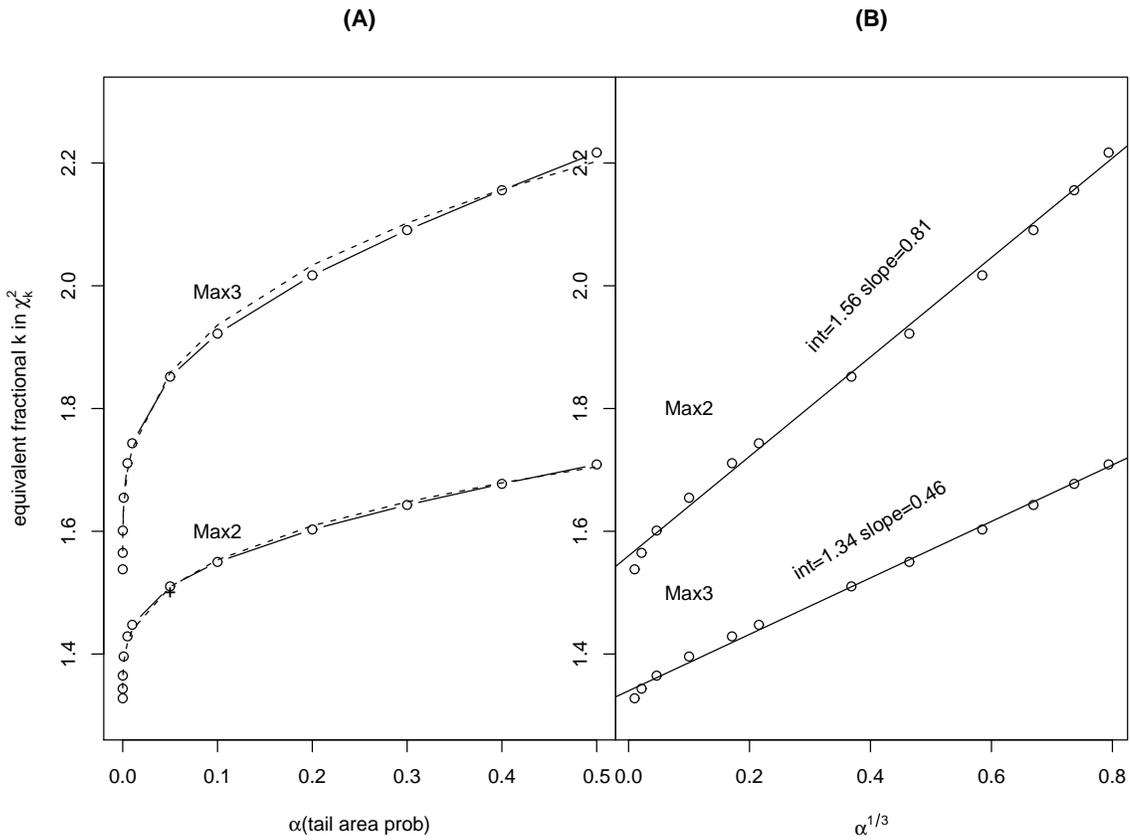, width=11cm}
  \end{turn}
\caption{
\label{fig5}
(A) The fractional degrees of freedom $k$ of $\chi^2_k$ that
is equivalent to Max2 and Max3 at tail area $\alpha$
as a function of $\alpha$.  
(B) $k$ is plotted versus $\alpha^{1/3}$.
}
\end{figure}

Fig.\ref{fig5}(A) shows the above-mentioned fractional $k$
value vs. the tail area probability $p_{Max2}$ or $p_{Max3}$. 
We are mostly interested in small tail area values, 
e.g. $p_{Max2}, p_{Max3} < 0.05$, in a test.  In this range, 
the equivalent fractional $k$ is constrained from above, e.g. smaller
than 1.5 (1.85) for Max2 (Max3). We also attempt to convert
the curve in Fig.\ref{fig5}(A) to a straight line by variable
transformation.  This can be accomplished by taking the 
cubic root of $p_{Max2}$ or $p_{Max3}$: in Fig.\ref{fig5}(B),
the fractional $k$ vs. $p_{Max2}^{1/3}$ or $p_{Max3}^{1/3}$ 
exhibit a reasonably good linear trend.

Besides fitting the tail area of Max2/Max3 by a fractional-$k$
$\chi^2_k$, one may also use a $\chi^2_k$ that has the same
average/mean as Max2 or Max3. We know that the average/mean of 
$\chi^2_k$ distribution is simply $k$, so this fractional dimension 
is very easy to determine. For example, in our simulation the means of Max2
and Max3 are 1.64 and 2.10 respectively. The corresponding
fractional-$k$ $\chi^2_k$'s that have the same mean would
be $\chi^2_{1.64}$ and $\chi^2_{2.10}$. Note that this fitting
of Max2/Max3 by $\chi^2_k$ is to fit the mean which receives
contribution from head as well as tail areas. It is not
surprising that the resulting $k$'s are different from those
that are based on tail areas only. Since the tail area is of
major concern in most statistical inferences, we regard
the definition of fractional $df$ from the tail area as more useful.

Figs.\ref{fig3}-\ref{fig5} all illustrate that a single fractional-$k$
$\chi^2_k$ cannot fit the Max2/Max3 distribution perfectly.
In particular, Fig.\ref{fig3}(C,D) shows that the
deviation between the two is directional: the matching
$\chi^2_k$ has a fatter tail than Max2/Max3 beyond the crossing
point.  One method to remove the systematic deviation in 
Fig.\ref{fig3}(C,D) is to use a linear function. For example, Fig.\ref{fig3}(D) 
show the result when the $-0.45+0.04  \chi^2_{k=1.7}$ linear trend is 
removed from the detrended QQ-plot of $\chi^2_{k=1.7}$ against 
Max3. It is equivalent to an approximation of sorted Max3 by 
$0.45+ 0.96 sort(\chi^2_{k=1.7})$.  Although it is not a perfect 
approximation, nor a unique one, the trend removal does reduce
the systematic deviation.

\section{Case-control genetic data} 

\indent

The result from the last section cannot be applied to the case-control
data directly because the individual test statistics in Eq.\ref{eq1}
are not independent, in particular for MAX3. There have been attempts 
to derive the null distribution of MAX3 by considering the joint 
distribution of $CAT(x=0)$, $CAT(x=0.5)$, and $CAT(x=1)$ \cite{gonzalez, qli}.
From Fig.3(B,D) and Fig.5(B), it is seen that we should not expect
a single $\chi^2_k$ with a fractional $k$ to fit the MAX3 distribution
perfectly.

The questions we asked for a real case-control data are:
(1) what are the approximate values of $k$ if a $\chi^2_k$
is forced to fit the tail area probability of MAX3?
(2) how good is our approximate distribution of Max3,
0.45+0.96 $\chi^2_{k=1.7}$, in fitting MAX3?  For answering these
questions, we use the case-control data for type 2 diabetes 
provided in \cite{sladek}.

The tail-area probability of MAX3 can be empirically obtained by
permutation: the affection status label of samples are randomly
shuffled, then the genotype counts are reconstructed. From
such a genotype count table, the MAX3 value can be determined.
Repeated calculation of MAX3 in label-shuffled dataset provides
a null distribution, and from which one can derive the tail-area 
probability. The $p_{MAX3}$ thus determined for the top SNPs
in \cite{sladek} is reproduced in Table 1. From the permutation-derived
$p_{MAX3}$, we find the best-fit $\chi^2_k$ that leads to the same
$p_{MAX3}$ value, and that fractional degrees of freedom $k$ is
listed in Table 1. A range of values of $k$ between 1.2 and 1.7, very
similar to the range used in Fig.3(D).

Next, we estimate the tail area probability of MAX3 by an approximate
formula discussed in the last section (and Fig.3(D)) for Max3.
Due to the difference of Max3 and MAX3, and the approximation nature
of the formula, we do not expect the derived tail area probabilities
to be exact. Surprisingly, from the result in Table 1, this approximation
actually leads to $p_{MAX3}$ that are similar to those obtained from
permutation in \cite{sladek}.

Permutation only provides a sampling of the null distribution, and
the finite number of replicates could be a source of error. 
Mimicking Fisher's exact test, which determines the tail area probability 
by counting the number of states in the tail area by combinatorics, 
we can also determine the exact value of $p_{MAX3}$ 
(J. Tian, C. Xu, H. Zhang, Y. Yang, paper in preperation).
This exact tail area probability is listed in the last column of
Table 1. Again, we see that the approximation of $p_{MAX3}$
based on fractional-$k$ chi-square distribution isn't far off
from the exact values.

\begin{table}
\begin{center}
\begin{tabular}{cccccc}
\hline
gene/SNP &MAX3 & $p_{MAX3}$ (permutation)$^{(1)}$ & $k^{(2)}$ & 
 $p_{MAX3}$ (by $k=1.7$ formula)$^{(3)}$ & $p_{MAX}$ (exact)$^{(4)}$ \\
\hline
TCF7L2/rs7900150 & 34.18437 & 2.1 $\times 10^{-8}$ & 1.676 & 1.36 $\times 10^{-8}$ & 1.29 $\times 10^{-8}$ \\
CAMTA1/rs1193179 & 25.71149& 6.3 $\times 10^{-7}$ & 1.213 & 1.17 $\times 10^{-6}$ & 1.00 $\times 10^{-6}$\\
CXCR4/rs932206 & 23.28708&  2.8 $\times 10^{-6}$ & 1.336 & 4.19 $\times 10^{-6}$ &  3.67 $\times 10^{-6}$ \\
ZNF615/rs1978717 & 23.11983&  4.9 $\times 10^{-6}$ & 1.595 &  4.57 $\times 10^{-6}$ & 4.01 $\times 10^{-6}$\\
HHEX/rs1111875 & 22.01918 & 8.6 $\times 10^{-6}$ & 1.597 &  8.17 $\times 10^{-6}$ & 7.82 $\times 10^{-6}$\\
LOC644419/rs282705 & 21.93485 & 9.0 $\times 10^{-6}$ & 1.598 & 8.54 $\times 10^{-6}$ &  6.27 $\times 10^{-6}$\\
\hline
\end{tabular}
\end{center}
\caption{
\textcolor{red}{
SNPs taken from the Table S4 of supplementary material of \cite{sladek}
with tail area probability (obtained from permutation) smaller than
$10^{-5}$, and if more than SNPs in a gene are significant at this
level, only one SNP is chosen here. The first two columns list the
gene/SNP name and the MAX3 value (based on the genotype counts given
in the supplementary material of \cite{sladek}). (1) values of tail area probability provided 
by \cite{sladek}; (2) the best fit of $k$ when the values in column 
``(1)" is used to fit a $\chi^2_k$ distribution; (3) estimation
of the tail area probability of MAX3 by the distribution (for Max3)
of 0.45+ 0.96$\chi^2_{k=1.7}$; (4) tail area probability of MAX3 
by the exact enumeration of all possible combinations.
}
}
\end{table}

\section{Discussion}

\indent

In this paper, we introduce two fractional parameter values 
for MAX test statistics:  (1) the fractional number of tests 
$d_{MAX}$ and (2) the fractional degree of freedom $k$ for the 
chi-square distribution that fits the Max null distribution.
The parameter $d_{MAX}$ has its counterparts in other 
fields, such as the effective number of parameters for model 
selection \cite{moody, mao, ye, spie},
effective number of genetic markers that are in linkage equilibrium
\cite{cheverud,nyholt}, effective
number of grid points required to represent a climate field
\cite{bretherton}, effective sample size in genetic
study for relatives \cite{yang}, etc. It was stated in
\cite{azzalini} that between the two extreme situations
of two tests being independent and being identical,
``an intermediate answer is to be anticipated". In one particular
situation, they actually have an example of 1.5 effective
number of tests (page 340 of \cite{azzalini}). 

There are two universal themes in these diverse studies: 
(1) Positive correlation causes the effective number to 
be smaller than the apparent number. 
This has several consequences, such as dimension reduction 
as a technique to simplify the dataset, correct ways 
for comparing statistical models by using the effective number 
of parameters to measure model complexity, etc. (2) As the 
effective number is determined from the real data, 
its value is most likely to be non-integer. Fractionality is the rule,
not an exception.

The Bonferroni correlation of $p$-value for multiple testing
is known to be conservative. The very reason that it is conservative
is because tests can be positively correlated, which is also
the cause for reduced values of effective number of tests.
Various attempts were made to take into account of correlation
among tests making a correction less conservative
\cite{simes,efron97,nino}.  Our simulation results show that the 
reduction of effective number of tests for MAX2 is very small, 
indicating that $CAT(x=0)$ and $CAT(x=1)$ are not strongly 
correlated. However, there is a large reduction in the effective 
number of tests for MAX3, and the multiple factor of 3 is not 
appropriate in Bonferroni correction for MAX3.

Non-integer degrees of freedom $k$ for $\chi^2_k$ is our second
fractional parameter, which had been encountered occasionally
in statistical literature (e.g., \cite{doornik}). The fact that
$df$ can be non-integer is not surprising by itself, but it is
more interesting to ask the question on whether it has any
geometric interpretation. Our case-control association analyses 
example may provide a hint, as there is a tangible link between 
the $k$ value and the size of area in the disease model space.

A disease model can be specified by 4 parameters (see Appendix),
but a projection from the 4-dimensional space to 2-dimensional
one is possible. Using a 2-by-3 genotype count table as a
realization of a disease model, Fig.\ref{fig6} shows two 
different ways to map a 2-by-3 genotype count table onto a 
two-dimensional plane.    The first, as shown in Fig.\ref{fig6}(A),
uses the case-control difference of Hardy-Weinberg disequilibrium
coefficients ($\delta_\epsilon$) and case-control difference of
allele frequency ($\delta_p$) (see Appendix) \cite{suh,li-cbc}.  The second, as shown
in Fig.\ref{fig6}(B), uses the odd ratio of the baseline and heterozygote
genotype ($OR_1$) and the odd ratio of the baseline and risk homozygote
genotype ($OR_2$) (see Appendix) \cite{ng}.

In the absence of constraints, randomly sampled disease model could
scatter within a bounded plane in Fig.\ref{fig6} (an outer bound for
Fig.\ref{fig6}(A) could be: $ -1 \le \delta_p \le 1$, $-1/2 \le \delta_\epsilon \le 1/2$),
whereas disease models in a given class are located in a more
restricted subspace, such as a line segment. We randomly sample 
dominant, recessive, multiplicative models and use them to 
generate dataset with 1000 case and 1000 control samples, these
generated genotype count tables are mapping to 2-dimensional
space in Fig.\ref{fig6}. In Fig.\ref{fig6}(A), multiplicative
models are located along the $y$-axis as this model does not lead
to Hardy-Weinberg disequilibrium; and recessive (dominant) models
are located in regions with positive (negative) $\delta_\epsilon$
values \cite{wittke, suh, li-cbc}. Similarly, in Fig.\ref{fig6}(B),
dominant models are located along the line with slope 1 ($OR_2=OR_1$),
multiplicative models are located in the line with slope 2
($\log(OR_2)/\log(OR_1)= 2$), and recessive models are on the
vertical line ($OR_1=1$ and arbitrary $OR_2$).

If we sort different test statistics according to their
corresponding degrees of freedom $k$ in $\chi^2_k$ for
the null distribution, the following order appears:
test on 2-by-3 genotype count table ($k=2$), MAX3 ($d \approx 1.57$), 
MAX2 ($d \approx 1.5$), $CAT(x=0.5)$ or $CAT(x=0)$ or $CAT(x=1)$ ($k=1$).
On the projected disease model space in Fig.\ref{fig6},
there is also a gradual narrowing of models for which these
tests are designed to detect: genotype count test targets
any models in the 2-dimensional space, MAX3 targets three
types of models represented by 3 line segments, MAX2
targets two types of models represented by 2 line segments,
and $CAT(x)$ targets only one line segments. 

In Fig.\ref{fig6}, the line segments for the three types of
disease models are somewhat blurred into wider areas, but
it is caused by random realization of datasets, rather than
a manifestation of a fractal geometry. However, the fractional
$df=k$ moves up from the integer value 1 to 1.5 and 1.57
when the number of line segments is increased. From this observation,
we do not believe fractional $k$ is related to a fractional
dimension of the underlying disease model subspace.

Even without a geometric interpretation, we may propose another
meaning for $k$: $k-1$ can be used to measure the level
of uncertainty in inferred disease model (mode of inheritance).
For $CAT(x)$, $k-1=0$, and a significant test result also
provides certain information concerning disease model. For
genotype test, $k-1=1$ and a significant test does not
tell us anything about the disease model. A significant MAX2
test result provides some information on disease model (for
example, that the true model is unlikely to be multiplicative),
whereas MAX3 offers even less information. If we consider
the detection of association signal and inference of disease
model as two independent tasks of a genetic association study,
then these two components are reminiscent of those studied
in the uncertainty principle in quantum physics \cite{heisenberg},
such as measuring the position and velocity of a particle
at the same time.

In conclusion, MAX test provides an interesting example
where two non-integer quantities can be defined and measured.
The effective number of tests to be maximized is more 
straightforward and has appeared in other applications as well.
The fractional-$k$ $\chi^2$ distribution for a test statistic 
is more intriguing, and seems to have a profound meaning concerning 
the test's ability to infer specific information. 
We have shown that a linear function of fractional-$k$ 
$\chi^2$ distribution approximates the true distribution of MAX
quite well. A hallmark of complex systems is its intermediate 
state between two extremes (order and disorder): a similarly 
intermediate state can also be described for fractional degrees 
of freedom in the MAX test which, in the genetic analysis context, 
sit between testing genetic association under completely 
specified and completely unknown disease models.

\section*{Acknowledgements} 

We would like to thank Jianan Tian for providing the {\sl R} code for
the exact calculation of tail area probability of MAX3, and Oliver
Clay for reading the first draft of the paper.

\newpage

\section*{Appendix: Basic notations and results for case-control genetic tests}

A case-control dataset consists of $N_1$ case samples and $N_0$
control samples whose genotype ($aa$ is the baseline homozygote,
$aA$ is the heterozygote, $AA$ is the risk homozygote) is known. The dataset can be
represented by a 2-by-3 genotype count table:

\begin{center}
\begin{tabular}{ccccc}
\hline
 &  aa & aA & AA & sample size   \\
\cline{2-4}
case(1) &  $N_{10}$ & $N_{11}$ & $N_{12}$ & $N_1=N_{1*}$ \\
control(0) &  $N_{00}$ & $N_{01}$ & $N_{02}$ & $N_0=N_{0*}$ \\
\hline
combined & $N_{*0}$ & $N_{*1}$ & $N_{*2}$ & $N=N_1+N_0$ \\
\hline
\end{tabular}
\end{center}

The above 2-by-3 genotype count table can be collapsed to several 2-by-2 tables. 
The following collapsing corresponds to a dominant model (the risk allele
$A$ ``dominates" allele ``a"):

\begin{center}
\begin{tabular}{ccc}
\hline
 &  aa & aA+AA  \\
\cline{2-3}
case(1) &  $N_{10}$ & $N_{11}$ + $N_{12}$ \\
control(0) &  $N_{00}$ & $N_{01}$ + $N_{02}$  \\
\hline
\end{tabular}
\end{center}
and the following collapsing corresponds to a recessive model
(only two copies of the risk allele $A$ present a disease risk):

\begin{center}
\begin{tabular}{ccc}
\hline
 &  aa+aA & AA  \\
\cline{2-3}
case(1) &  $N_{10}$ + $N_{11}$ &  $N_{12}$ \\
control(0) &  $N_{00}$ + $N_{01}$ & $N_{02}$  \\
\hline
\end{tabular}
\end{center}
 
From a 2-by-2 table, Pearson's chi-square test statistic $X^2$
is of the form of $\sum_{row, col} (O_{row, col}-E_{row, col})^2/E_{row, col}$ 
where $O_{row,col}$ is the observed (genotype) count is a table cell
indexed by ``row" and "column", and $E_{row,col}$ is the
expected count. The expected count is equal to the product of
the row margin $O_{row, *}= \sum_{col} O_{row, col}$ and the column margin
$O_{*, col}= \sum_{row} O_{row, col}$. It can be shown that $X^2$ is
the product of squared matrix determinant and total sample size
divided by the product of 4 row and column margins (e.g., \cite{suh}). 
For example, for the recessive model, $X^2$ is:
\begin{eqnarray}
 D &=&  (N_{10}+N_{11}) N_{02} - (N_{00}+N_{01}) N_{12} \nonumber \\
 X^2_{REC} &=& \frac{ D^2 N}{ (N_{*0}+N_{*1})N_{*2} N_1 N_0}
\end{eqnarray}
Under the null hypothesis (by chance alone), $X^2$ follows the
$\chi^2_{k=1}$ distribution (chi-square distribution with one degree
of freedom). $X^2_{DOM}$ can be calculated similarly.

The Cochran-Armitage trend ($CAT$) test is defined after each genotype
is assigned a score. Most assignment of the genotype score could be
equivalent to a score of $\{ x_i \} \equiv (0, x, 1)$, 
i.e., the score for the baseline homozygote is fixed at 0, 
that for the risk homozygote is fixed at 1, and that for the 
heterozygote is a parameter $x$. The $CAT$ test statistic at $x$
is defined as (\cite{sasieni,slager,zheng06}:
\begin{equation}
CAT(x) = \frac{N_{1*} N_{0*} }{ N}
\frac{ \left(  \sum_{j=0}^2 x_j (N_{1j}/N_{1*}-  N_{0j}/N_{0*}) \right)^2}
{ \left( \sum_{j=0}^2 x_j^2 N_{*j}/N - (\sum_{j=0}^2 x_j N_{*j}/N)^2
\right)}
\nonumber
\end{equation}
It can be shown that $CAT(x=0)$ is equal to $X^2_{REC}$ and
$CAT(x=1)$ is equal to $X^2_{DOM}$. Under the null hypothesis,
$CAT(x)$ at each fixed $x$ value follows the $\chi^2_{k=1}$ distribution.

A disease model of a bi-allelic disease locus can be specified by 4 parameters. 
One is the allele frequency ($p \equiv p_A$) and the other three 
characterize the susceptibility of the disease under each genotype:
$(f_0, f_1, f_2) \equiv (P(disease|aa), P(disease|Aa), P(disease|AA))$.
The latter three parameters can be replaced by the following three parameters:
relative genotype risk for heterozygote: $\lambda_1 \equiv f_1/f_0$,
that for the risk homozygote, $\lambda_2 \equiv f_2/f_0$, and
disease prevalence $K= f_0[ (1-p)^2 + \lambda_1 2p(1-p) +\lambda_2 p^2]$.
Either a ($p_A, f_0, f_1, f_2$) value or a ($p_A, \lambda_1, \lambda_2, K$) value
uniquely determines a disease model.

There are several ideas in reducing the number of parameters of
a disease model from 4 to 2 ``major" parameters. One suggestion
\cite{suh} is to use the allele frequency difference in case and in control
group $\delta_p \equiv p_A(case) - p_A(control)= p_1 - p_0$, and Hardy-Weinberg
disequilibrium coefficient difference in the two groups
$\delta_\epsilon \equiv \epsilon(case)- \epsilon(control) = 
\epsilon_1 - \epsilon_0$. The Hardy-Weinberg disequilibrium coefficient
$\epsilon$ measures the deviation from Hardy-Weinberg equilibrium \cite{weir},
such that the three genotype frequencies can be written as ($(1-p)^2+\epsilon$,
$2p(1-p) - 2\epsilon$, $p^2+ \epsilon$). The motivation for this
parameterization is that $\delta_p$ is directly related to the
case-control association signal, and $\delta_\epsilon$ is 
strongly correlated with the disease model.

The group-specific allele frequency and Hardy-Weinberg disequilibrium
coefficient can be determined from the 4 parameters $p, \lambda_1, \lambda_2, K$ 
\cite{wittke}, and their differences can be
determined as well \cite{li-cbc}:
\begin{eqnarray}
\delta_p \equiv
p_1 - p_0 &=& \frac{f_0( p^2 \lambda_2 + p(1-p) \lambda_1)}{K}
- \frac{ p^2 (1-f_0\lambda_2)+p(1-p)(1-f_0\lambda_1)}{ 1-K}
 \nonumber \\
\delta_\epsilon \equiv
\epsilon_1 - \epsilon_0 &=& \frac{f_0^2p^2(1-p)^2 (\lambda_2-\lambda_1^2)}{K^2}
- \frac{f_0 p^2(1-p)^2 (2\lambda_1-1-\lambda_2-f_0\lambda_1^2+f_0\lambda_2)}
{(1-K)^2}. \nonumber 
\end{eqnarray}
Given a 2-by-3 genotype table, these two parameters can be estimated by:
\begin{eqnarray}
\label{eq:delta-data}
\hat{\delta}_p &=& \hat{p}_1-\hat{p}_0 =  \frac{N_{12}+N_{11}/2}{N_1} -
\frac{N_{02}+ N_{01}/2}{N_0}
 \nonumber \\
\hat{\delta}_\epsilon &=& \hat{\epsilon}_1 - \hat{\epsilon}_0 =
\frac{N_{12}}{N_1} - \left( \frac{N_{12}+N_{11}/2}{N_1}  \right)^2
- \frac{N_{02}}{N_0} + \left( \frac{N_{02}+ N_{01}/2}{N_0} \right)^2.
 \nonumber
\end{eqnarray}

\begin{figure}[th]
  \begin{turn}{-90}
   \epsfig{file=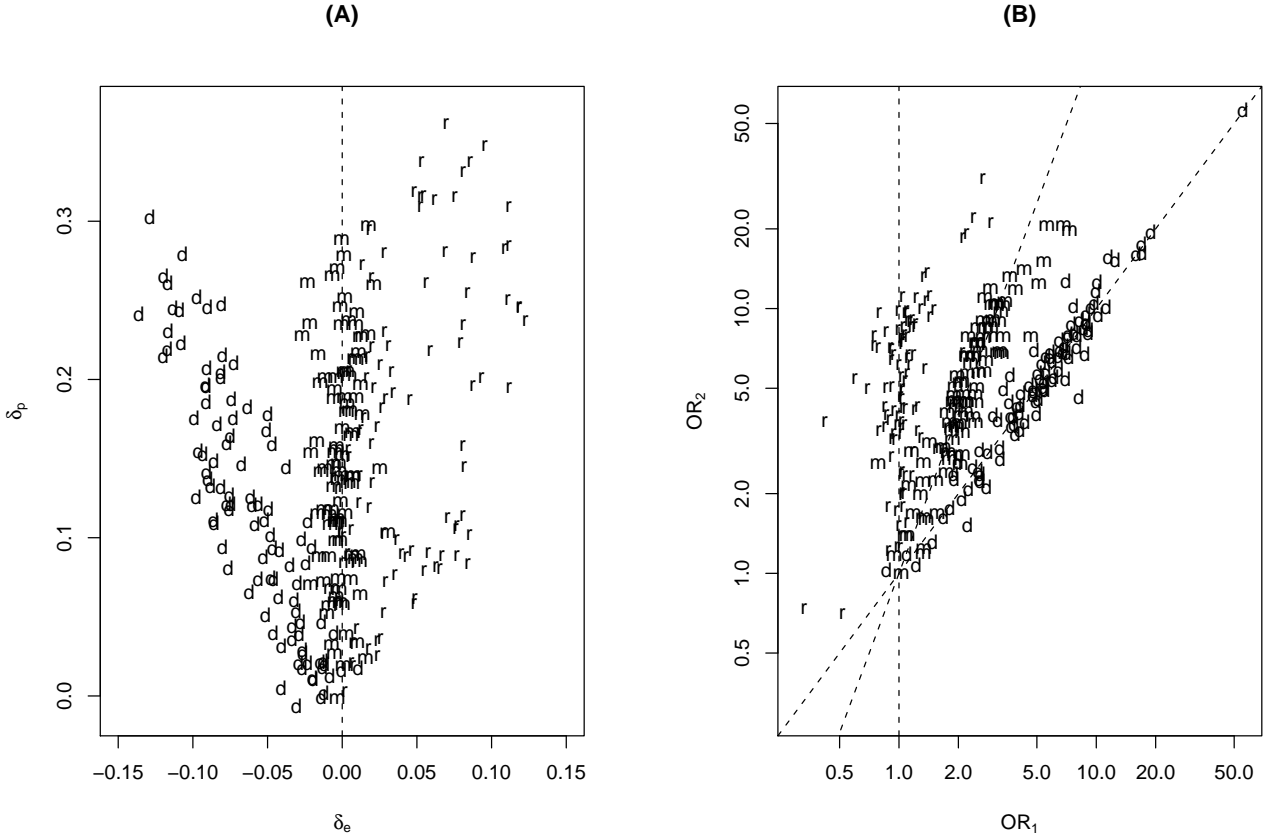, width=11cm}
  \end{turn}
\caption{
\label{fig6}
Simulation of 100 case-control datasets each for three classes of
models (dominant, multiplicative, recessive). Each point represents
a genotype count table for 1000 cases and 1000 controls. The allele
frequency is randomly chosen from (0.1-0.9); disease prevalence
is sampled from the normal distribution with a random mean, and 
standard deviation of 1/10 of the mean; the $\lambda_2$ genotype
relative risk is randomly chosen between (1.1-10); $\lambda_1$
is equal to $\lambda_2$, $\sqrt{\lambda_2}$, and 1 for dominant,
multiplicative, and recessive models.
(A) The location of simulated datasets in the $\delta_\epsilon$-$\delta_p$ 
parameter space, where $\delta_\epsilon$ is the case-control difference of Hardy-Weinberg
disequilibrium coefficients and $\delta_p$ is the case-control
difference of allele frequencies. The symbols ``d", ``m", ``r"
represent dominant, multiplicative, and recessive models, respectively.
(B) The location of the same simulated case-control datasets in the
$OR_1$-$OR_2$ space (both $x$ and $y$-axis are in log scale), 
where $OR_1$ is the odd-ratio of heterozygote genotype vs. baseline 
homozygote genotype, and $OR_2$ is the odd-ratio of risk homozygote genotype
vs. baseline homozygote genotype.
}
\end{figure}

Another idea in selecting two major parameters in the disease model
is to ignore $p$ and $K$, and focus only on $\lambda_1$ and $\lambda_2$.
These two parameters can be estimated by the two odd-ratios from
subtables consisting of one baseline column and another risk column:
\begin{eqnarray}
\hat{\lambda_1}= OR_1 &=& \frac{ N_{11} N_{00}}{ N_{10} N_{01}} \nonumber \\
\hat{\lambda_2}= OR_2 &=& \frac{ N_{12} N_{00}}{ N_{10} N_{02}}. \nonumber 
\end{eqnarray}
Fig.\ref{fig6}(A) and (B) illustrate these two ideas of
a two-dimensional disease model space.

\newpage

\newpage

\newpage

\newpage

\end{document}